\documentstyle[preprint,aps]{revtex}

\begin{document}
\title{Magnetic field-induced mixing of hyperfine states of Cs 6$^{2}$P$_{3/2}$
level observed with a sub-micron vapor cell}
\author{Aram Papoyan, David Sarkisyan}
\address{Institute for Physical Research, NAS of Armenia, Ashtarak-2, 378410 Armenia}
\author{Kaspars Blush, Marcis Auzinsh}
\address{Department of Physics, University of Latvia, 19 Rainis blvd., Riga\\
LV-1586, Latvia}
\author{Daniel Bloch, Martial Ducloy}
\address{Labaratoire de Physique des Lasers, UMR 7538 du CNRS, Institut Galilee,\\
Universit\'{e} Paris-Nord, 93430 Villetaneuse, France\\
(version 3.0)}
\maketitle

\begin{abstract}
The fluorescence spectra of a sub-micron atomic cesium vapor layer
observable under resonant excitation on D2 line have been studied in the
presence of an external magnetic field. Substantial changes in amplitudes
and frequency positions of the individual (resolved) hyperfine transitions
have been recorded in moderate magnetic fields (up to $\sim $ 50 Gauss).
These features are caused by mixing of the hyperfine states of the upper
level resulting from comparable values of the hyperfine splitting of the 6$%
^{2}$P$_{3/2}$ manifold and Larmor frequencies of the magnetic sublevels.
The results of simulation show a good agreement with the \vspace{0in}%
experimental spectra. Possible application of the results for high spatial
resolution magnetometry is discussed. {\em PACS number(s)}: ???
\end{abstract}

\section{Introduction}

It is well known that application of an external magnetic field modifies
absorptive and dispersive properties of an atomic system. As magnetic field
strength starts to increase from zero value, the degeneracy of atomic states
lifts, and magnetic sublevels diverge. The linear shift of these sublevels
takes place until the $B$- field values for which the Zeeman splitting
becomes comparable to the hyperfine splitting of atomic levels. In this
regime, the wave functions of the hyperfine manifold begin to mix with each
other. This mixing results in dual consequences. First, the shift of
magnetic sublevels is no more linear on $B$- field (and of different
behavior for different individual sublevels). Second, probabilities of the
optical transitions between the magnetic sublevels of the lower and upper
atomic states differ from unperturbed values \cite{trem1}. For the first
excited levels of Rb and Cs, the atoms of particular interest for
spectroscopic studies, the linear Zeeman regime ceases at $B<100$ Gauss.
Thus, crossing of magnetic sublevels for 5$^{2}$P$_{3/2}$ manifold of $^{85}$%
Rb occurs at $B\approx 4$ Gauss \cite{auz4}. Since magnetic splitting in
these conditions is incomparably smaller than Doppler broadening, direct
observation of hyperfine sublevels mixing effects in absorption and
fluorescence spectra for vapor media becomes rather complicated. That is why
the studies were done either for higher magnetic fields \cite{trem1}, or by
non-direct methods like selective reflection spectroscopy \cite{weis1,pap1}
at lower magnetic fields. Also the alignment to orientation conversion
observation can be used as a very sensitive tool to study level mixing due
to different perturbations (see \cite{auz3,auz7} and references therein)
including atoms with hyperfine structure in a magnetic field \cite{auz4}.

Development of extremely thin vapor cells (ETC) of a sub-micron thickness 
\cite{sark1} has opened new possibilities for high resolution spectroscopy
of vapor (gaseous) media, allowing one to directly record sub-Doppler lines
in the fluorescence and transmission spectra. In particular, as is shown in 
\cite{sark1}, all the six hyperfine transitions of the Cs atomic D2 line are
completely resolved in the fluorescence spectrum (Fig.\ref{levels} shows
hyperfine levels and transitions relevant for this line). Moreover, it
should be noted that the amplitude ratios for the individual hyperfine
transitions are practically independent of exciting radiation intensity. The
latter indicates that the optical depopulation pumping on the non-cycling
transitions $F_{g}$ $=3,4\rightarrow F_{e}$ $=3,4$ doesn't establish even at
the laser intensities of $I_{L}\sim 50$ mW/cm$^{2}$ for the sub-micron vapor
column length. This property is of importance for quantitative spectroscopic
and magneto-optical measurements.

Absorption and emission processes in the sub-micron-thick dilute vapor layer
are strongly affected by anisotropic geometrical factor. For the atoms
flying with mean thermal velocity normally to the window surface (that is
''Doppler-sensitive'' direction), the wall-to-wall flight time is only $\sim 
$1 ns, i.e., much less than inverse natural width of the optical transition.
Fast atoms simply do not have time to absorb, and especially to emit before
getting de-excited on the dielectric surface. As a result, the contribution
of slow longitudinal velocity atoms in the spectra becomes predominant,
giving rise to Dicke-type sub-Doppler features. And since the use of the ETC
allows to substantially diminish the Doppler broadening of spectral lines,
the influence of the magnetic field on individual hyperfine components can
be revealed at weaker magnetic fields as compared with well-known
experiments on upper-state level mixing and crossing in Doppler-broadened
gases. Hence, the gradual ${\bf I}-{\bf J}$ decoupling as a magnetic field
increases can be studied in detail.

Up to now, magnetic field-induced effects in thin cells of radiation
wavelength-scale thickness have been studied only theoretically \cite{zam1}.
In this study, peculiarities of Faraday rotation and magnetic circular
dichroism have been considered.

In present work, we have studied the influence of magnetic field on
fluorescence spectra of a sub-micron Cs vapor layer. The choice of
fluorescence rather than absorption studies was justified by better spectral
resolution of the individual hyperfine lines \cite{sark1}.

\section{Experimental}

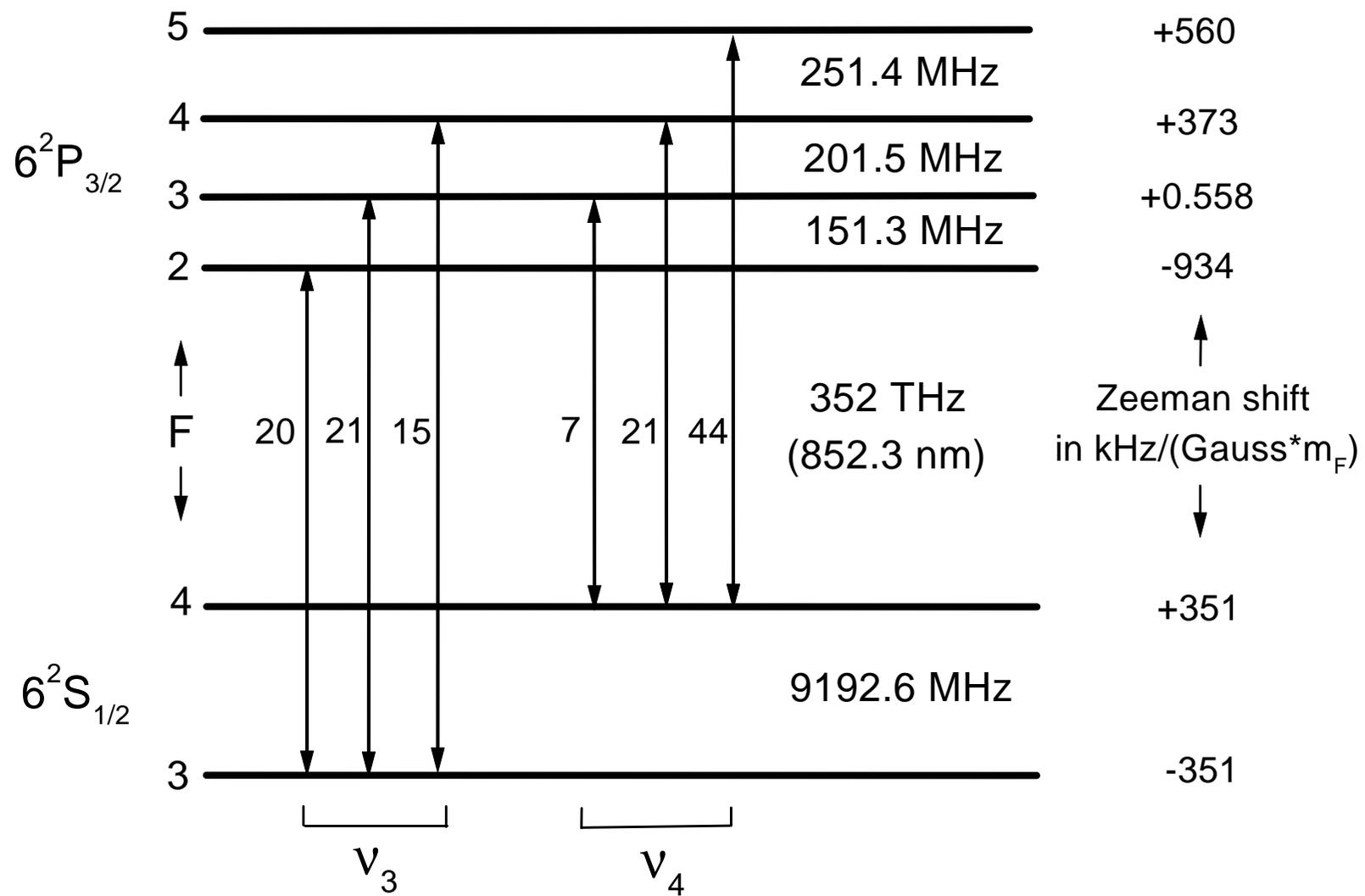
\begin{figure}[tbp]
\caption{Geometrical configuration of the experiment. Laser radiation
propagation ${\bf k}$ is along {\it z} axis; arrows show various
polarization cases (linear, circular left- and right-handed). Magnetic field
is applied along {\it x}, {\it y}, or {\it z} axis. Cs ETC: extremely thin
(sub-micron) cell containing Cs vapor.}
\label{config}
\end{figure}

Schematic drawing of the measurement configuration is shown in Fig.\ref
{config}. The radiation beam ($\oslash 3$ mm) of the $\lambda =$ $852$ nm,
20 MHz-linewidth single-frequency {\it cw} laser diode was directed at
normal incidence onto the ETC of $\sim $300 nm thickness with the side arm
containing cesium. The cell was placed in the 3 pairs of mutually
perpendicular Helmholtz coils providing possibility to cancel the ambient
magnetic field and to apply homogeneous magnetic field in arbitrary
direction. The laser frequency was linearly scanned in the region of $\nu
_{3}$ or $\nu _{4}$ transitions of Cs atom (Fig.\ref{levels}). A
Glan-Thomson prism was used to purify original (linear) radiation
polarization of the laser diode; to produce a circular polarization, a $%
\lambda /4$ plate was utilized. A photodiode followed by an operation
amplifier was placed at 90$%
{{}^\circ}%
$ to the laser propagation direction to detect the fluorescence signal
emerging through one of two side openings of the cell oven. The photodiode
collected emission within $\sim $ 0.1 srad solid angle.

Intensity of the fluorescence emission (with no spectral and polarization
analysis) from the vapor layer excited by linearly and circularly polarized
radiation was recorded versus the laser radiation frequency, for various
directions and magnitudes of the external magnetic field. For the case of
longitudinal magnetic field (${\bf B}//{\bf k}$), notations of right- and
left-handed circular (elementary) polarizations are $\sigma ^{+}$ and $%
\sigma ^{-}$, respectively. As one could expect, switching from $\sigma ^{+}$%
($\sigma ^{-}$) to $\sigma ^{-}$($\sigma ^{+}$) by adjusting the $\lambda /4$
plate keeping invariable direction of the magnetic field $+B$ in this case
was completely identical to switching magnetic field from $+B$ to $-B$
keeping invariable polarization helicity $\sigma ^{+}$($\sigma ^{-}$).
Indeed, the recorded signals in these two cases were the same, and for
technical convenience, the second option has been exploited for the regular
measurements. Over 80 experimental spectra have been recorded with the $B$%
-field strength ranged from 0 to 55 Gauss (the maximum field produced by the
set of Helmholtz coils we have used), with the step of 11 Gauss. In fact,
some influence of the magnetic field is clearly observable already at $B\sim
10$ Gauss, but to avoid cumbersome graphical presentation, we will restrict
ourselves by the limiting cases of $B=0$ and $B=55$ Gauss (in the latter
case the effect of the $B$-field is the most visible).

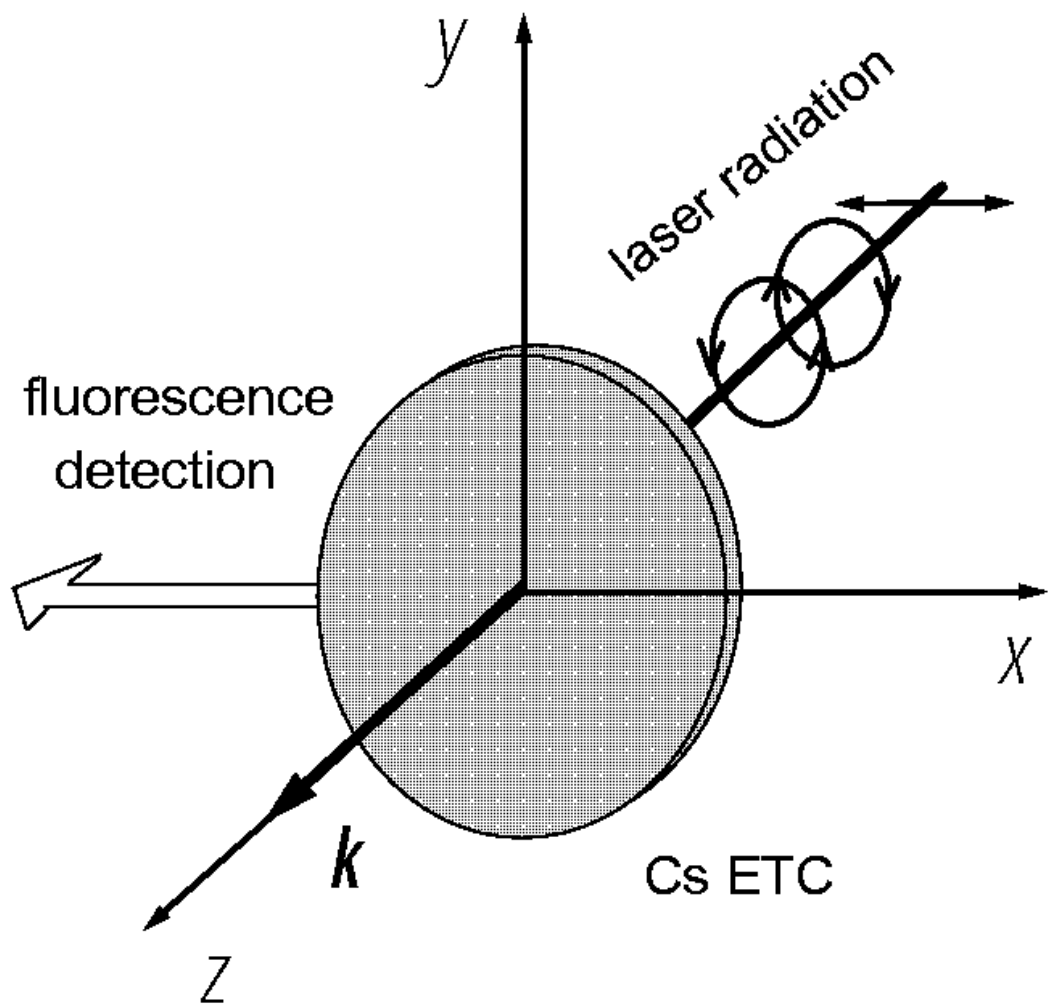
\begin{figure}[tbp]
\caption{Experimental ({\it scatter} {\it circles}) and theoretical ({\it %
lines}) fluorescence spectra of $\protect\nu _{4}$: $F_{g}=4\rightarrow
F_{e}=(2),3,4,5$ ({\it a}) and $\protect\nu _{3}$: $F_{g}=3\rightarrow
F_{e}=2,3,4,(5)$ ({\it b}) transitions without magnetic field ({\it upper row%
}) and with $B_{z}=55$ Gauss magnetic field ({\it lower row}). Incident
radiation polarization: linear ({\it left column}), circular right-handed (%
{\it middle column}), and left-handed ({\it right column}). The frequency
increases rightwards; the arbitrary units of the vertical (intensity) axis
are the same for all the graphs.}
\label{Bz}
\end{figure}

Scatter circles in Figs.\ref{Bz}-\ref{By} show the examples of the recorded
fluorescence spectra. The orientations of the $B$-field are: $B=B_{z}$ (Fig.%
\ref{Bz}), $B=B_{x}$ (Fig.\ref{Bx}), and $B=B_{y}$ (Fig.\ref{By}). Figs.\ref
{Bz}-\ref{By},{\it a} represent the results for $\nu _{3}$ transitions ($%
F_{g}=3\rightarrow F_{e}=2,3,4$), and Figs.\ref{Bz}-\ref{By},{\it b} show
the results for $\nu _{4}$ transitions ($F_{g}=4\rightarrow F_{e}=3,4,5$)%
\footnote{%
These notations are valid for $B=0$. As ${\bf I}-{\bf J}$ decoupling starts
to develop in the $B$-field, the levels that initially are not accessible
due to dipole transiton selection rules, start to contribute to optical
transitions. Hence, it is more correct to write $F_{g}=3\rightarrow
F_{e}=2,3,4,(5)$ for $\nu _{3}$ and $F_{g}=4\rightarrow F_{e}=(2),3,4,5$ for 
$\nu _{4}$.}. In all the figures, the upper row of graphs are $B=0$ spectra,
and the lower row graphs are $B=55$ Gauss spectra. Notations for every
column indicate the applied laser radiation polarization and $B$-field
orientation. In fact, the same $B=0$ spectra with linear and circular laser
radiation polarization are presented in the upper rows of all these figures,
and they are repeated just to stress the modifications induced by a magnetic
field in every particular case (for visual comparison).\ The vertical
(intensity) scale is the same for all the graphs; the horizontal (frequency)
scale is 200 MHz/div, the frequency rises rightwards. Saturated absorption
signal from an auxiliary 1.5 cm-long Cs cell has been used as a frequency
marker (reference). The temperature of the side arm reservoir of the cell
(the latter defined the saturated vapor pressure) was kept at $T_{r}=105$ 
${{}^\circ}$%
C throughout the measurements. Corresponding number density of Cs atoms is $%
N_{Cs}=2\times 10^{13}$ cm$^{-3}$, and the width of collisional
(homogeneous) broadening of the spectral lines is $\approx $ $1.8$ MHz, i.e.
nearly 3 times smaller as compared with the natural line width. The Doppler
(inhomogeneous) line width at this temperature regime is $\approx $ $440$
MHz.

In agreement with nearly linear intensity dependence of the fluorescence
reported in \cite{sark1}, the shape of fluorescence spectrum does not
noticeably depend on laser radiation intensity, at least in the range of $%
I_{L}=1-50$ mW/cm$^{2}$. All the spectra presented in graphs were recorded
with $I_{L}=40$ mW/cm$^{2}$. For $I_{L}$ $<1$ mW/cm$^{2}$, the fluorescence
signal is too weak to be recorded. On the other hand, focussing of the laser
beam is required to get higher laser intensities $I_{L}$ $>50$ mW/cm$^{2}$,
but in this case the interaction time (time of flight) changes
simultaneously; focussing also decreases the number of atoms contributing to
the fluorescence signal. The spectra remain invariable under reversal of
direction of the laser frequency scanning. Substantial difference in spectra
for right-handed ($\sigma ^{+}$)\ and left-handed ($\sigma ^{-}$) circular
polarizations of the laser radiation is observed only for longitudinal
direction of the $B$-field ($B=B_{z}$, Fig.\ref{Bz}). For the case of
transverse magnetic field (Figs.\ref{Bx},\ref{By}), the helicity of light is
of no relevance, and the corresponding spectra are identical. Moreover, the
spectra with circularly-polarized excitation for $B=B_{x}$ (Fig.\ref{Bx})
and $B=B_{y}$ (Fig.\ref{By}) are practically identical. These observations
are well understandable from symmetry reasons.

As we can see from Figs.\ref{Bz}-\ref{By}, the individual magnetic
subtransitions $m_{Fg}$ $\rightarrow $ $m_{Fe}$ of the same hyperfine
transition $F_{g}$ $\rightarrow $ $F_{e}$ are not resolved in the conditions
of present experiment. Indeed, in the case of complete resolution, one could
expect to see 18 magnetic transitions for the $\nu _{3}$ line and 24
magnetic transitions for the $\nu _{4}$ line for each of elementary
polarizations: $\pi $ (linear, ${\bf B}//{\bf E}$), $\sigma ^{+}$ and $%
\sigma ^{-}$ (circular, ${\bf B}//{\bf k}$), and up to 88 magnetic
transitions for more complex $B$-field orientation cases. The results of
experiment can be rather interpreted as $B$-field orientation- and
magnitude-dependent frequency shift and variation of amplitude of the $F_{g}$
$\rightarrow $ $F_{e}$ hyperfine transitions.

We would like to draw special attention to the results depicted in \ref{Bz},
with dramatic difference between the fluorescence spectra for the $\sigma
^{+}$ and $\sigma ^{-}$ polarizations. This non-intuitive behavior reflects
magnetic circular dichroism of the Cs atoms (i.e., different absorption
coefficients for $\sigma ^{+}$ and $\sigma ^{-}$ light in the magnetic
field). The most surprising result is that the $F_{g}=3\rightarrow F_{e}=4$
(Fig.\ref{Bz}{\it a}, $\sigma ^{-}$ polarization) and $F_{g}=4\rightarrow
F_{e}=3$ (Fig.\ref{Bz}{\it b}, $\sigma ^{+}$ polarization) are found to be
completely suppressed at rather low magnetic fields.

\section{Model}

We will use the following model to simulate the spectra of the Cs atomic
film without a magnetic field and in the magnetic field. First, we will
assume that the absorption rate $\Gamma _{p}$ is small in comparison with
the relaxation rates in ground and excited states, denoted by $\gamma $ and $%
\Gamma $ respectively: $\Gamma _{p}<\gamma ,\Gamma $. This assumption is
well justified by the observation of fluorescence spectrum independence of
the laser intensity. Second, as far as laser line is rather broad $\Delta
\nu _{L}\approx 20$ MHz, the laser radiation spectral line that excites
atoms is broader than the homogeneous width of the atomic transition. The
radiation width of the transition is $\Gamma _{rad}=1/(2\pi \tau )\approx
5.2 $ MHz, where $\tau \approx 30.5$ ns is lifetime of Cs in 6$^{2}$P$_{3/2}$
state \cite{theo1}.

As we will see further, the conditions for laser excitation of atoms differ
substantially when magnetic field is switched off and when it is applied.
Let us start with the analysis of excitation of atoms in zero magnetic field.

\subsection{Zero magnetic field signal}

A convenient way to describe excited state atom optical excitation in the
broad line approximation is by means of a quantum density matrix $%
f_{mm^{\prime }}$\cite{auz5}. We will consider an atom that absorbs laser
light polarized in the direction characterized by light electric field
vector ${\bf E}_{exc}$. In the limit of weak absorption, light does not
affect atoms in the ground state. All the ground state magnetic sublevels
remain equally populated, and no coherences between magnetic sublevels in
the ground state are created. In this situation the density matrix of the
excited state can be calculated as \cite{auz6}:

\begin{equation}
f_{mm^{\prime }}=\frac{\Gamma _{p}}{\Gamma }\sum_{\mu }\left\langle
F_{e}m\right| \widehat{{\bf E}}_{exc}^{\ast }{\bf \cdot }\widehat{{\bf d}}%
\left| F_{g}\mu \right\rangle \left\langle F_{e}m^{\prime }\right| \widehat{%
{\bf E}}_{exc}^{\ast }{\bf \cdot }\widehat{{\bf d}}\left| F_{g}\mu
\right\rangle ^{\ast }.  \label{eq1}
\end{equation}
Transition takes place from the ground state hyperfine level $F_{g}$ to the
excited state hyperfine level $F_{e}$. Magnetic quantum numbers of the
ground state hyperfine level $F_{g}$ are denoted by $\mu $ and magnetic
quantum numbers of the excited state hyperfine level $F_{e}$ by $m$ and $%
m^{\prime }$. Matrix elements of the dot product can be written as \cite
{var1} 
\begin{equation}
\left\langle F_{e}m\right| \widehat{{\bf E}}^{\ast }\cdot \widehat{{\bf d}}%
\left| F_{g}\mu \right\rangle =\sum_{q}\left( E^{q}\right) ^{\ast
}\left\langle F_{e}m\right| d^{q}\left| F_{g}\mu \right\rangle ,  \label{eq2}
\end{equation}
where $E^{q}$ are cyclic components of light polarization vector. These
components, which in general case are complex numbers, are connected with
more familiar Cartesian components as follows \cite{auz5,var1}: 
\begin{eqnarray}
E^{+1} &=&-\frac{1}{\sqrt{2}}(E_{x}-iE_{y}),  \nonumber \\
E^{0} &=&E_{z},  \label{eq3} \\
E^{-1} &=&\frac{1}{\sqrt{2}}(E_{x}+iE_{y}).  \nonumber
\end{eqnarray}
The remaining matrix element $\left\langle F_{e}m\right| d^{q}\left|
F_{g}\mu \right\rangle $ can be written in an explicit form by means of
Wiegner--Eckart theorem \cite{var1,sob1,zar1}: 
\begin{equation}
\left\langle F_{e}m\right| d^{q}\left| F_{g}\mu \right\rangle
=(-1)^{F_{e}-m}\left( 
\begin{array}{ccc}
F_{e} & 1 & F_{g} \\ 
-m & q & \mu
\end{array}
\right) \left\langle F_{e}\right\| d\left\| F_{g}\right\rangle ,  \label{eq4}
\end{equation}
where $\left( 
\begin{array}{ccc}
\circ & \circ & \circ \\ 
\circ & \circ & \circ
\end{array}
\right) $ is 3-$jm$ symbol. To calculate still remaining reduced matrix
element $\left\langle F_{e}\right\| d\left\| F_{g}\right\rangle $, we will
assume the following scheme of formation of hyperfine level angular momentum
in ground and excited states: 
\begin{equation}
{\bf J}_{g}+{\bf I}={\bf F}_{g};\qquad {\bf J}_{e}+{\bf I}={\bf F}_{e},
\label{eq41}
\end{equation}
where ${\bf J}_{g}$ and ${\bf J}_{e}$ denote total electronic angular
momentum of an atom in the ground and excited states (including electron
spin ${\bf S}$), and ${\bf I}$ is the nuclear spin angular momentum. This
scheme allows to write the angular part of wave functions of atom in the
following form: 
\[
\left| F_{g}\right\rangle =\left| \left( J_{g}I\right) F_{g}\right\rangle
\qquad \left| F_{e}\right\rangle =\left| \left( J_{e}I\right)
F_{e}\right\rangle , 
\]
and taking into account that optical transition dipole moment operator acts
only upon the electronic angular momentum and does not act upon the nuclear
spin angular momentum, we will get the following expression for the
remaining reduced matrix element \cite{sob1,zar1}:

\begin{equation}
\left\langle F_{e}\right\| d\left\| F_{g}\right\rangle
=(-1)^{J_{g}+I+F_{e}+1}\sqrt{(2F_{e}+1)(2F_{g}+1)}\left\{ 
\begin{array}{ccc}
J_{g} & F_{g} & I \\ 
F_{e} & J_{e} & 1
\end{array}
\right\} \left\langle J_{e}\right\| d\left\| J_{g}\right\rangle ,
\label{eq5}
\end{equation}
where $\left\{ 
\begin{array}{ccc}
\circ & \circ & \circ \\ 
\circ & \circ & \circ
\end{array}
\right\} $ is 6$j$ symbol. These last formulas allow to calculate excited
state density matrix for different ground and excited state hyperfine
levels, rather easy.

Finally, the fluorescence signal for each allowed hyperfine transition from
the excited state hyperfine level $F_{e}$ to the final hyperfine level $%
F_{g} $ (which may or may not coincide with the level from which absorption
took place) can be calculated according to \cite{auz5}: 
\begin{equation}
I=\stackrel{\symbol{126}}{I}_{o}\sum_{mm^{^{\prime }}\mu }\left\langle
F_{e}m\right| \widehat{{\bf E}}_{obs}^{\star }\widehat{{\bf d}}\left|
F_{g}\mu \right\rangle \left\langle F_{e}m^{^{\prime }}\right| \widehat{{\bf %
E}}_{obs}^{\star }\widehat{{\bf d}}\left| F_{g}\mu \right\rangle ^{\star
}f_{m^{\prime }m},  \label{eq6}
\end{equation}
where ${\bf E}_{obs}$ denotes the polarization of the light to which
detector is sensitive. In the case of fluorescence, the general approach to
calculation of dipole transition matrix elements entering Eq.(\ref{eq6}) is
the same as above described method to calculate matrix elements entering
expression Eq.(\ref{eq1}) for excited state density matrix calculation. If
we measure the total intensity of fluorescence, on transitions to both
ground state hyperfine levels without discrimination for a specific light
polarization, than we simply must take a sum of two fluorescence components
with orthogonal polarizations observed in a definite direction and sum of
two spectral transitions to the both ground state hyperfine levels with $%
F_{g}=3$ and $F_{g}=4$. Such a spectrum simulated for linearly and
circularly polarized excitation is given on Fig.\ref{illu}{\it a,b}. Left
and right graphs of the upper rows refer to the absorption of linearly and
circularly polarized light, respectively, from the ground state hyperfine
level $F_{g}=4$ (case {\it a}) and $F_{g}=3$ (case {\it b}). The dashed line
represents the sum of three Lorentz type curves centered around the
respected hyperfine component frequency. Width of this Lorentz type spectral
line is chosen to be equal to $55$ MHz. This is a width which gives signals
that fit well to experimentally observed ones. For the experimental signal
detected in this work, this width includes laser linewidth, absorption
linewidth formed by homogeneous linewidth (natural and collisional), and
residual Doppler broadening. Of course, when we have a homogeneous linewidth
combined with inhomogeneous broadening, the resulting line shape should be a
Voigt contour, but just to make model as simple as one could, Lorentz type
line shape for each individual peak was chosen to produce Fig.\ref{illu}.
This assumes that remaining Doppler broadening is smaller than the
homogeneous linewidth. One can see that the relative intensities of spectral
components in this simulation differ from relative line strength which can
be calculated as \cite{auz2} 
\begin{eqnarray}
W_{F_{g}\leftarrow \rightarrow F_{e}}
&=&(2F_{e}+1)(2F_{g}+1)(2J_{e}+1)(2J_{g}+1)  \nonumber \\
&&\times \left\{ 
\begin{array}{ccc}
J_{g} & F_{g} & I \\ 
F_{e} & J_{e} & 1
\end{array}
\right\} ^{2}\left\{ 
\begin{array}{ccc}
L_{g} & J_{g} & S \\ 
J_{e} & L_{e} & 1
\end{array}
\right\} ^{2}  \label{eq_n6}
\end{eqnarray}
where $L_{g}$ and $L_{e}$ are electron orbital momenta for the ground and
excited state.

This difference has an obvious explanation. As far as we are using polarized
light to excite atoms, the excited state spatial angular momenta
distribution of atoms is anisotropic. This anisotropy is different for
different types of transitions and depends on quantum numbers $F_{g}$ and $%
F_{e}$. It is especially sensitive to the quantity $\Delta F=F_{g}-F_{e}$.
As a result, radiation emission from the excited state hyperfine levels
towards a specific direction in a space differs, among other factors like
transition strength, also because of this anisotropy. For more detailed
explanation of this effect, see for example \cite{the1,auz1}.

\subsection{Signal with a magnetic field}

When an external magnetic field is applied, the Hamilton operator for the
atom in a magnetic field can be written as

\begin{equation}
\widehat{H}=\widehat{H}_{0}+\widehat{H}_{HFS}-{\bf \mu }_{J}{\bf \cdot B-\mu 
}_{I}{\bf \cdot B.}  \label{eq7}
\end{equation}
where $\widehat{H}_{0}$ is a Hamiltonian operator of unperturbed atom, $%
\widehat{H}_{HFS}$ represents hyperfine interaction. The remaining two terms
represent interaction of the electronic magnetic moment ${\bf \mu }_{J}$ of
atom and the nucleus magnetic moment ${\bf \mu }_{I}$ with the external
magnetic field ${\bf B}$. These magnetic moments are connected with the
respective electronic and spin angular moments{\bf \ }${\bf J}$ and ${\bf I}$
of the atom: 
\begin{equation}
{\bf \mu }_{J}{\bf =}\frac{g_{J}\mu _{B}}{\hbar }{\bf J,\qquad \mu }_{I}{\bf %
=}\frac{g_{I}\mu _{0}}{\hbar }{\bf I,}  \label{eq8}
\end{equation}
where $\mu _{B}$ and $\mu _{0}$ are the Bohr and nuclear magnetons
respectively, and $g_{J}$, $g_{I}$ are electrotonic and nuclear Land\'{e}
factors. The action of magnetic field on the atom has two closely related
effects. First, magnetic sublevels of the hyperfine levels are mixed in the
magnetic field: 
\begin{equation}
\left| \gamma _{k}m\right\rangle
=\sum_{F_{e}=J_{e}-I}^{F_{e}=J_{e}+I}C_{kF_{e}}^{\left( e\right)
}(B,m)\left| F_{e},m\right\rangle ,\qquad \left| \eta _{j}\mu \right\rangle
=\sum_{F_{i}=J_{i}-I}^{F_{i}=J_{i}+I}C_{jF_{i}}^{\left( i\right) }(B,\mu
)\left| F_{i},\mu \right\rangle ,  \label{eq9}
\end{equation}
where $C_{kF_{e}}^{\left( e\right) }(B,m)$ and $C_{jF_{i}}^{\left( i\right)
}(B,\mu )$ are mixing coefficients depending on the field strength and
magnetic quantum number ($m$ or $\mu $). The second effect is deviation of
the Zeeman magnetic sublevel splitting in the magnetic field for each
hyperfine level form the linear one. It means that the additional energy of
the magnetic sublevel obtained in the magnetic field is not any more
linearly proportional to the field strength. In general case, new atomic
states $\left| \gamma _{k}m\right\rangle $ and $\left| \eta _{j}\mu
\right\rangle $ in the magnetic field are linear combinations of all initial
hyperfine levels (4 in the case of Cs atoms in 6$^{2}$P$_{3/2}$ state and 2
in case of Cs atom in the 6$^{2}$S$_{1/2}$ state). As it is seen from Eq.(%
\ref{eq9}), the hyperfine angular momentum quantum number $F$ ceases to be a
good quantum number in the magnetic field, but magnetic quantum numbers $m$
and $\mu $ are still good quantum numbers. This reflects the symmetry of the
perturbation imposed by the magnetic field and means that only hyperfine
sublevels with the same magnetic quantum numbers are mixed by the magnetic
field.

The mixing coefficients $C_{kF_{e}}^{\left( e\right) }(B,m)$ and $%
C_{jF_{i}}^{\left( i\right) }(B,\mu )$ of the hyperfine states in the
magnetic field and energies of these levels $^{\gamma _{k}}E_{m}$, $^{\eta
_{j}}E_{\mu }$ can be found as eigenvectors and eigenvalues of the Hamilton
matrix (\ref{eq7}). In Fig.\ref{enmag}, the energy levels obtained by the
Hamilton matrix diagonalization for Cs atom in the excited 6$^{2}$P$_{3/2}$
state in the magnetic field are presented.

For Cs atoms in the ground state hyperfine level, the splitting exceeds 9
GHz, which is large in comparison to the magnetic sublevel energies obtained
in the magnetic field. As a result, ground state energy levels in the
magnetic field can with very good approximation be represented by the linear
Zeeman effect. Namely $^{\eta _{j}}E_{\mu }=g_{\eta _{j}}\mu _{B}B\mu /\hbar 
$, where $g_{\eta _{j}}$ is the Land\'{e} factor of the respective hyperfine
level. For very weakly mixed levels they still can be represented with the
hyperfine quantum number $F_{g}$. For Cs atoms in 6$^{2}$S$_{1/2}$ state we
have $g_{\eta _{j}}=-1/4$ for $F_{g}=3$, and $g_{\eta _{j}}=1/4$ for $%
F_{g}=4 $. In the case of mixing of only two hyperfine levels, an analytical
formula can be derived for mixing coefficients and for level energies in the
magnetic field, similar to Breit Rabi formula, see for example \cite{alex1}.

The excited state density matrix created by the laser light in the magnetic
field can be written as (see for example \cite{auz4})

\begin{equation}
^{kl}f_{mm^{\prime }}=\frac{\widetilde{\Gamma }_{p}}{\Gamma +i^{kl}\Delta
\omega _{mm^{\prime }}}\sum_{j\mu }\left\langle \gamma _{k}m\right| \widehat{%
{\bf E}}_{exc}^{\ast }{\bf \cdot }\widehat{{\bf D}}\left| \eta _{j}\mu
\right\rangle \left\langle \gamma _{l}m^{\prime }\right| \widehat{{\bf E}}%
_{exc}^{\ast }{\bf \cdot }\widehat{{\bf D}}\left| \eta _{j}\mu \right\rangle
^{\ast }.  \label{eq10}
\end{equation}
$^{kl}\Delta \omega _{mm^{\prime }}=(^{\gamma _{k}}E_{m}-^{\gamma
_{l}}E_{m^{\prime }})/\hbar $ is the energy splitting of the magnetic
sublevels $m$ and $m^{\prime }$ belonging to the excited state levels $k$
and $l$. Magnetic quantum numbers of the ground state level $\eta _{j}$ are
denoted by $\mu $ and magnetic quantum numbers of the excited state level $%
\gamma _{k,l}$ by $m$ and $m^{\prime }$. In this last expression it is
assumed that two magnetic sublevels of the excited state, that initially
belonged to two different hyperfine levels, can have the same energy and can
be excited simultaneously and coherently at some specific magnetic field
strength. This means that nonzero field level crossing signals \cite{auz4}
are included in this model. At the same time, for practical calculations
performed in this work this is of no relevance, since no nonzero magnetic
field level crossings take place at the values $B\leq 55$ Gauss used in the
experiment.

In our particular simulation, we assume that when we scan the laser
frequency in the manifold of splitted in the external magnetic field
magnetic sublevels, only those transitions that are in an exact resonance
with the laser field are excited.\ So at each radiation frequency, a
specific for this frequency density matrix is calculated. Of course, density
matrix is dependent also on the magnetic field strength, which determines
magnetic sublevel splitting and wave function mixing and as a result,
transition probabilities between magnetic sublevels.

In general case, the intensity of the fluorescence with a specific
polarization ${\bf E}_{obs}$ in a transition between excited $\gamma _{k}$
and final $\eta _{j}$ state in the magnetic field can be calculated
according to \cite{auz4}:

\begin{equation}
I\left( {\bf E}_{f}\right) =I_{0}\sum_{mm^{\prime }\mu
}\sum_{klj}\left\langle \gamma _{k}m\right| \widehat{{\bf E}}_{obs}^{\ast }%
{\bf \cdot }\widehat{{\bf D}}\left| \eta _{j}\mu \right\rangle \left\langle
\gamma _{l}m^{\prime }\right| \widehat{{\bf E}}_{obs}^{\ast }{\bf \cdot }%
\widehat{{\bf D}}\left| \eta _{j}\mu \right\rangle ^{\ast kl}f_{mm^{\prime
}}.  \label{eq4a}
\end{equation}
Final state of the transition may or may not coincide with the atomic ground
state from which the absorption started. In lower row graphs of Fig.\ref
{illu}{\it a},{\it b}, the fluorescence spectra (line positions and relative
intensities) are shown as vertical bars, for the case of excitation from $%
F_{g}=4$ ({\it a})\ and $F_{g}=3$ ({\it b}),\ with linear and circular light
polarizations. The dashed lines again show the resulting spectra supposing
that each component has a Lorentz shape line with $55$ MHz width. The
results for $B=55$ Gauss in Fig.\ref{illu} represent graphical illustration
of the calculation approach for the case of elementary polarizations ($\pi $%
, $B=B_{x}$, and $\sigma ^{+}$, $B=B_{z}$). For other geometrical
configurations, the situation is more complex to be presented by vertical
bars, since two or even three elementary polarization components may be
involved.

This approach to calculate spectra in the magnetic field and in the absence
of the field was used in the simulation of experimentally observed signals.
The simulation was done for all the experimentally recorded spectra shown in
Figs.\ref{Bz}-\ref{By}, the results are presented by solid lines. One can
see rather good agreement between the theoretical and experimental results.
Some discrepancy observable in the graphs may be attributed to the following
factors. First, the simulation model implies Lorentz profile, rather than
Voigt one, for individual transitions, and this simplification may introduce
some spectral distortions. Second, the cell geometry (narrow gap between the
thick windows) imposes additional (polarization sensitive) asymmetry on the
spatial distribution of the emission. The fluorescence emitted under
noticeable range of angles around 90$%
{{}^\circ}%
$ may be guided towards the photodiode by [multiple] grazing reflections
from the external/internal faces of the window. Third, the recorded spectra
may be affected by anisotropic elastic collisions of polarized atoms with
the cell walls.

\section{Conclusion}

The resonance fluorescence of Cs vapor layer of sub-micron thickness has
been studied in the presence of moderate ($\sim $ 50 Gauss) external
magnetic field, under excitation of atoms by the laser radiation tuned to
the frequency region of D2 line. The spectra recorded for various reciprocal
orientations of magnetic field and laser polarization (linear, circular)
exhibit substantial frequency shift, modification of peak amplitude and line
shape for individual hyperfine transitions. These changes originate from
mixing of the hyperfine levels of the upper state 6$^{2}$P$_{3/2}$ and
become observable because of intrinsic sub-Doppler nature of atomic signal
in an extremely thin cell. The simulation has been performed by means of a
quantum density matrix in the broad line approximation for a weak absorption
regime, taking into account mixing of magnetic sublevels of the hyperfine
levels and their non-linear shift in magnetic field. The results of
simulation show good agreement with the experimental results, which
indicates that other possible effects (anisotropy in long-range atom-surface
interaction, coherent and propagation effects, etc.) have no practical
contribution in the conditions of our study.

Zeeman splitting of hyperfine sublevels in vapor media for moderate magnetic
field has been observed in selective reflection \cite{pap1} and saturated
absorption \cite{schm1}\ spectroscopy. Nevertheless, direct recording of
fluorescence (and possibly absorption) spectra using extremely thin cells is
a more straightforward method, since the signal in this case is not
influenced by contribution from other unavoidable processes (dispersion and
non-linear effects, crossover resonances, etc.). Sub-Doppler resolution that
is a necessary condition for observation of Zeeman splitting is satisfied
also for atomic beams, as well as for cold atoms. In these cases, however,
cumbersome optical setups are required. Moreover, application of a magnetic
field may perturb initial properties of atomic system (the case of cold
atoms in magneto-optical traps).

The results of present work can be used for high spatial resolution
magnetometry. Indeed, the interaction of laser radiation with atomic system
responsible for magnetic sensitivity takes place in the sub-micron gap
between the cell windows. Note that also the transverse dimensions of the
interaction region can be essentially reduced (to the micrometer range) by
means of focussing the laser beam. Extremely small sensor size of
corresponding magnetometer device may allow the fine mapping of magnetic
field spatial distribution that can be important for many applications, in
particular for testing high-gradient magnetic fields. The magnetic field
sensitivity can be noticeably enhanced by use of Rb D2 line where the
smaller splitting of the upper state (5$^{2}$P$_{3/2}$) will cause
recordable level mixing effects at much lower $B$- fields ($\sim $ 1 Gauss).
It should be noted that though suggested device will have much lower
sensitivity than other optical magnetometer schemes (in particular, ones
based on nonlinear Faraday and Hanle effects), there are several advantages
that justify its possible application, namely: i) unique spatial resolution,
ii) simplicity of optical setup, iii) extended upper limit of measured $B$-
fields.

\begin{description}
\item 
\begin{description}
\item  {\bf Acknowledgments}. This work was supported, in part, by grants
\#00-378 and \#00-381 of the Ministry of Economics of Armenia, as well as by
ANSEF grant No. PS18-01. The authors are grateful to A. Sarkisyan for his
valuable contribution to fabrication of the ETC, and to Yu. Malakyan for
stimulating discussions.
\end{description}
\end{description}

\begin{figure}[tbp]
\caption{Experimental ({\it scatter} {\it circles}) and theoretical ({\it %
lines}) fluorescence spectra of $\protect\nu _{4}$: $F_{g}=4\rightarrow
F_{e}=(2),3,4,5$ ({\it a}) and $\protect\nu _{3}$: $F_{g}=3\rightarrow
F_{e}=2,3,4,(5)$ ({\it b}) transitions without magnetic field ({\it upper row%
}) and with $B_{x}=55$ Gauss magnetic field ({\it lower row}). Incident
radiation polarization: linear ({\it left column}), and circular ({\it right
column}). The frequency increases rightwards; the arbitrary units of the
vertical (intensity) axis are the same for all the graphs.}
\label{Bx}
\end{figure}

\begin{figure}[tbp]
\caption{Experimental ({\it scatter} {\it circles}) and theoretical ({\it %
lines}) fluorescence spectra of $\protect\nu _{4}$: $F_{g}=4\rightarrow
F_{e}=(2),3,4,5$ ({\it a}) and $\protect\nu _{3}$: $F_{g}=3\rightarrow
F_{e}=2,3,4,(5)$ ({\it b}) transitions without magnetic field ({\it upper row%
}) and with $B_{y}=55$ Gauss magnetic field ({\it lower row}). Incident
radiation polarization: linear ({\it left column}), and circular ({\it right
column}). The frequency increases rightwards; the arbitrary units of the
vertical (intensity) axis are the same for all the graphs.}
\label{By}
\end{figure}
\begin{figure}[tbp]
\caption{Illustration of the fluorescence spectrum simulation for $\protect%
\nu _{4}$ ({\it a}) and $\protect\nu _{3}$ ({\it b}) transitions. {\it %
Vertical bars} show the relative strength of the individual fluorescence
components ($\left| F_{e}m\right\rangle \rightarrow $ $\left| F_{g}\protect%
\mu \right\rangle $ for $B=0$,{\it \ upper row}; and $\left| \protect\gamma
_{k}m\right\rangle \rightarrow $ $\left| \protect\eta _{j}\protect\mu
\right\rangle $ for $B=55$ Gauss, {\it lower row}). {\it Dashed lines}
represent the sum of these (broadened) components.}
\label{illu}
\end{figure}
\begin{figure}[tbp]
\caption{Splitting and shift of the hyperfine states of the 6$^{2}$P$_{3/2}$
level in the magnetic field.}
\label{enmag}
\end{figure}

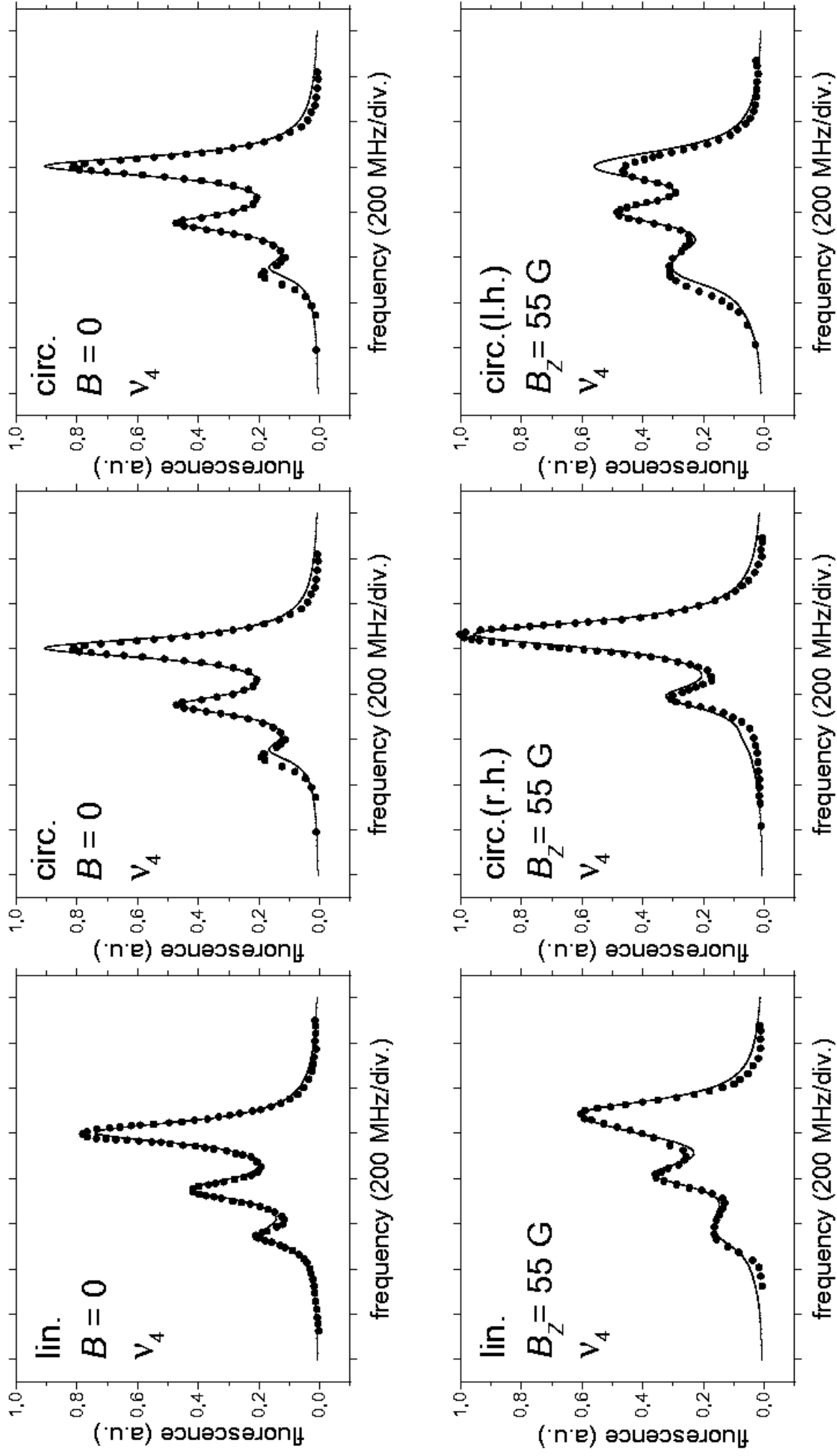
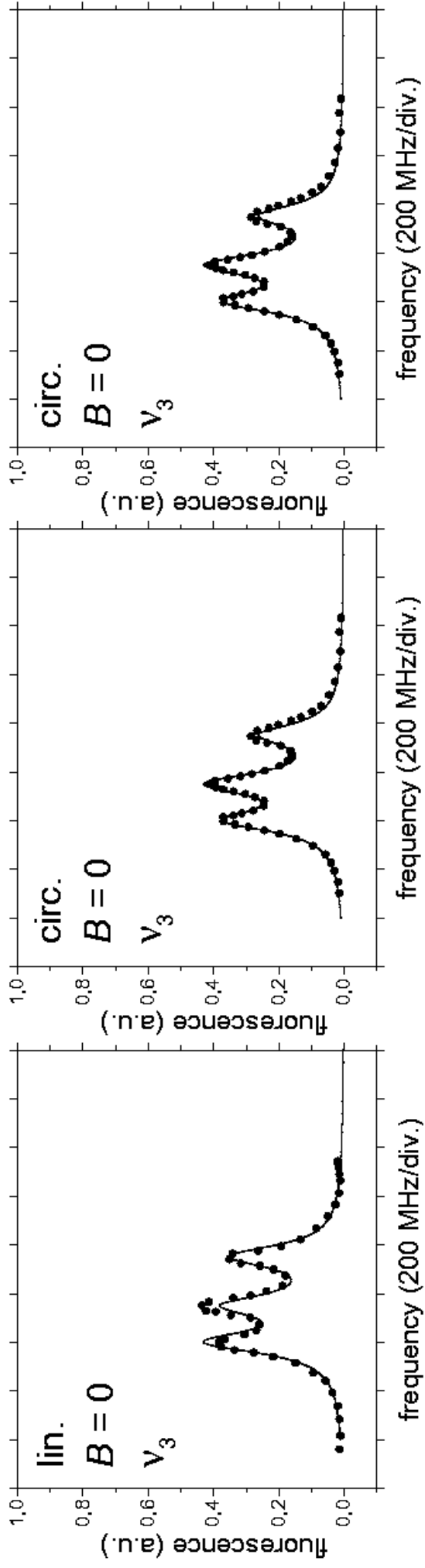
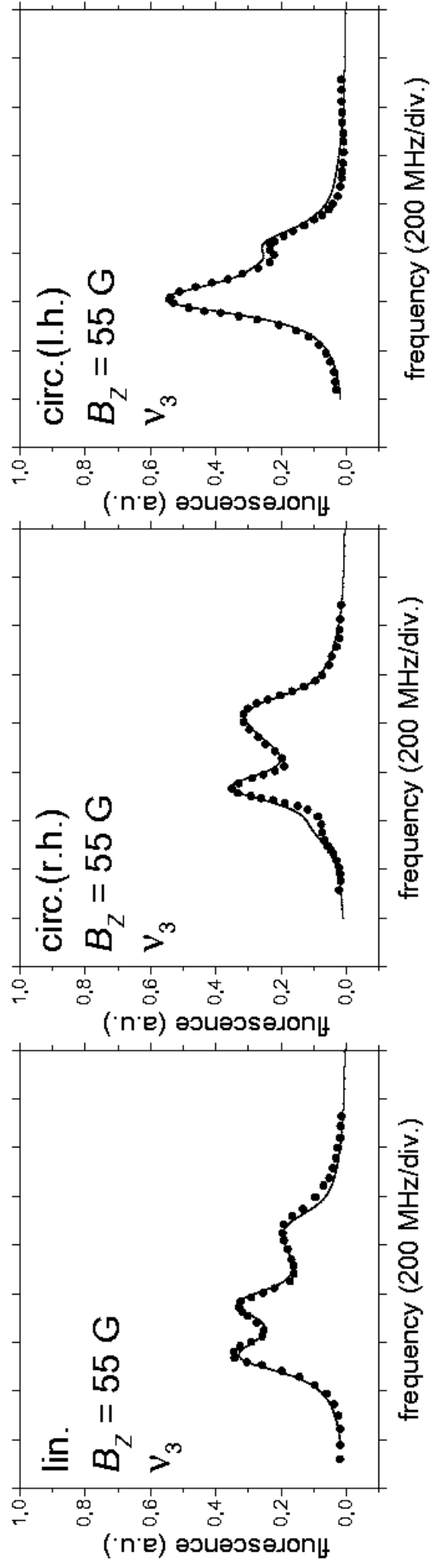
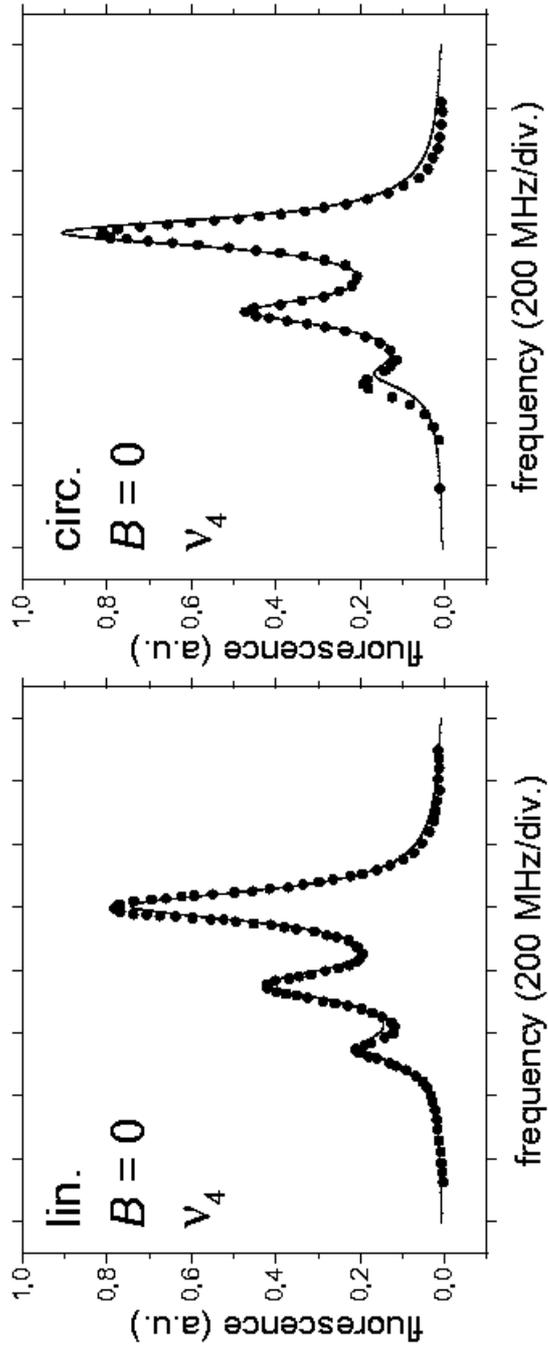
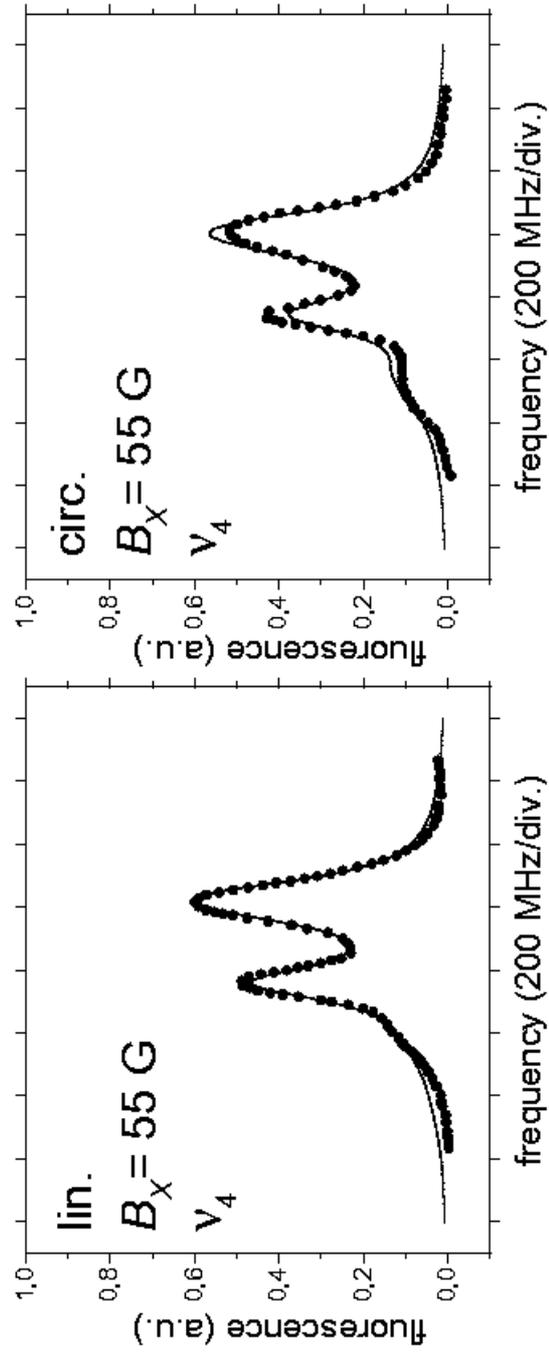
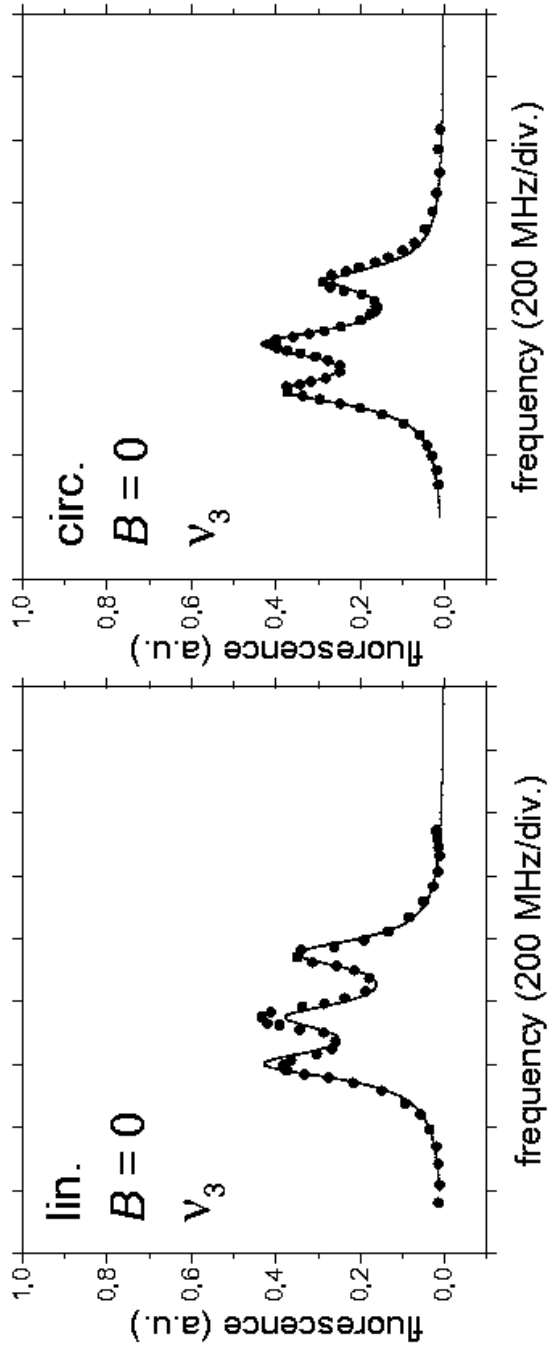
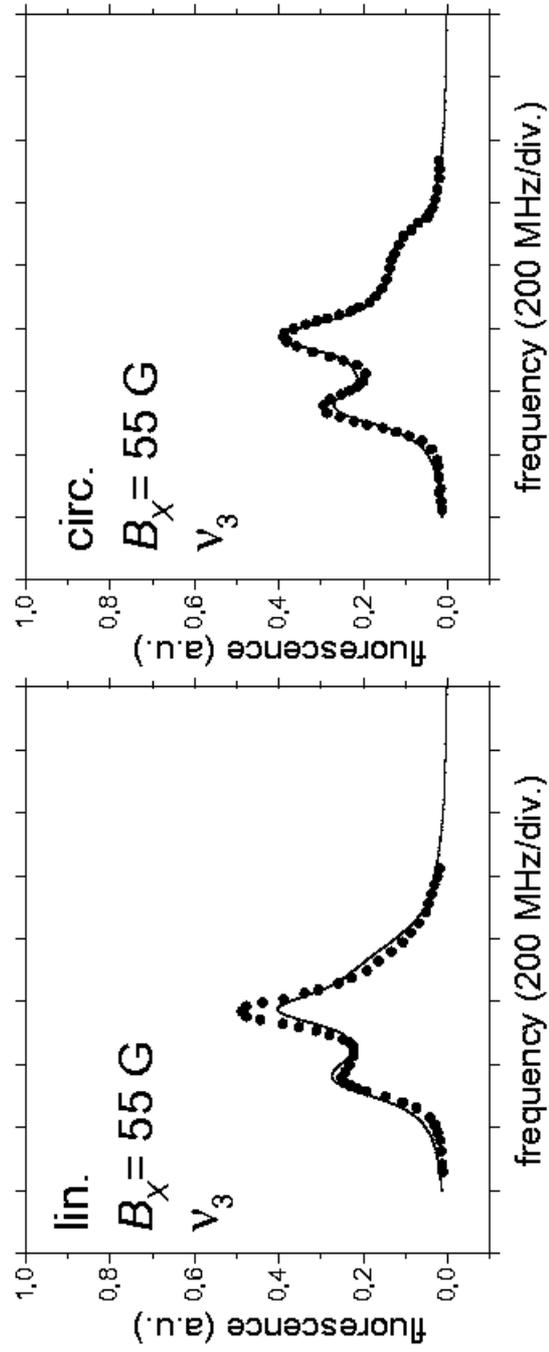
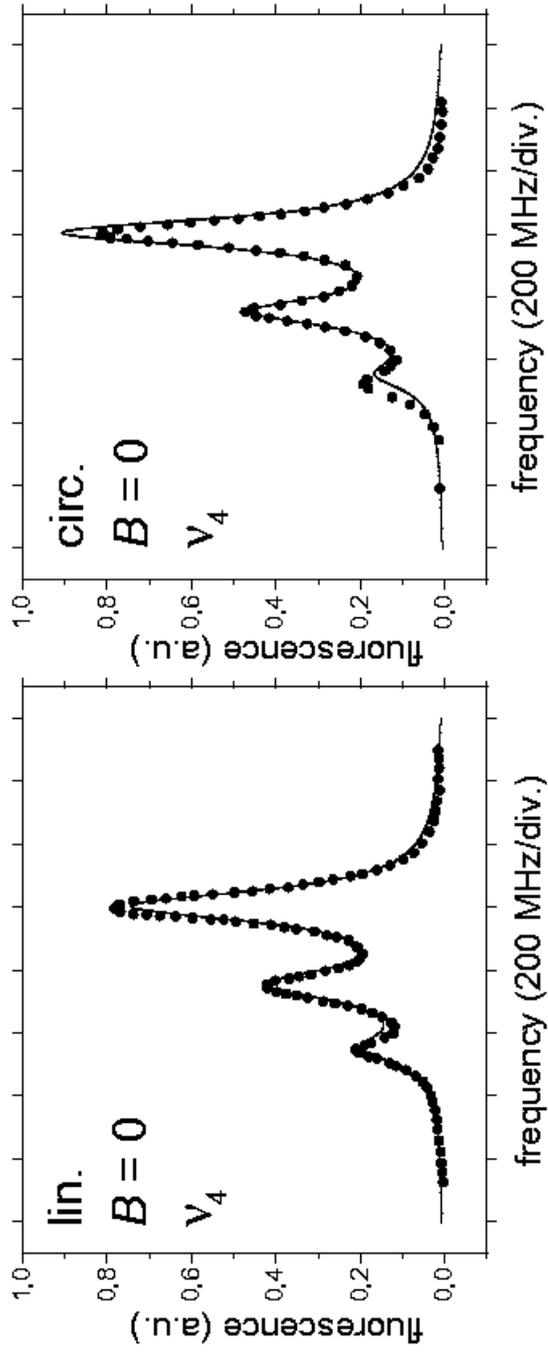
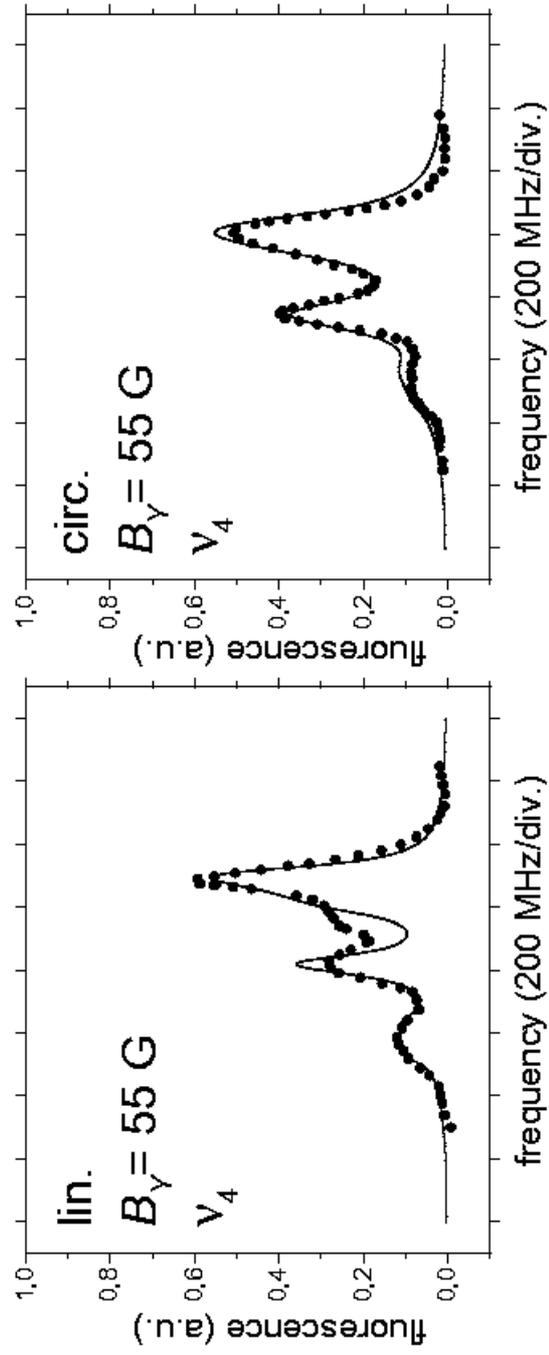
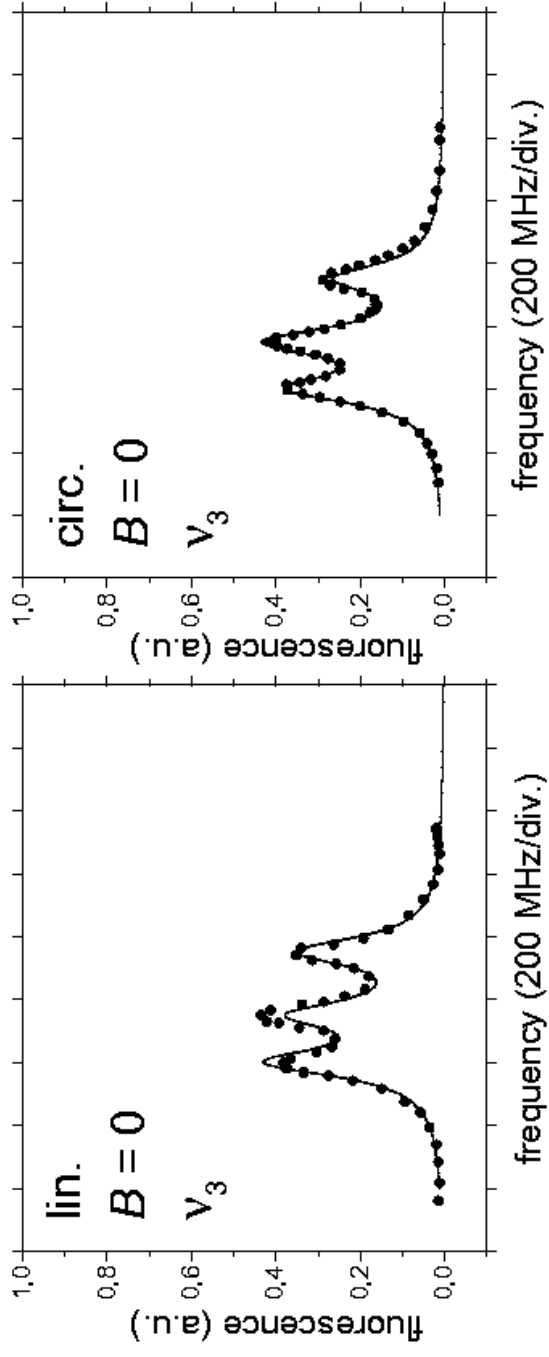
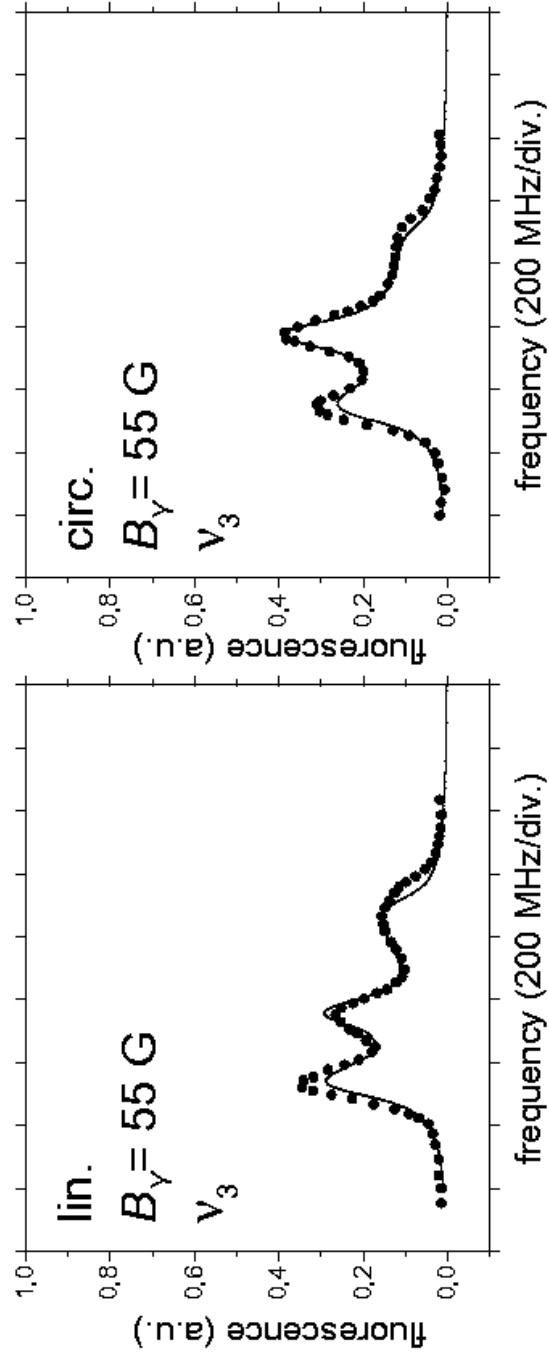
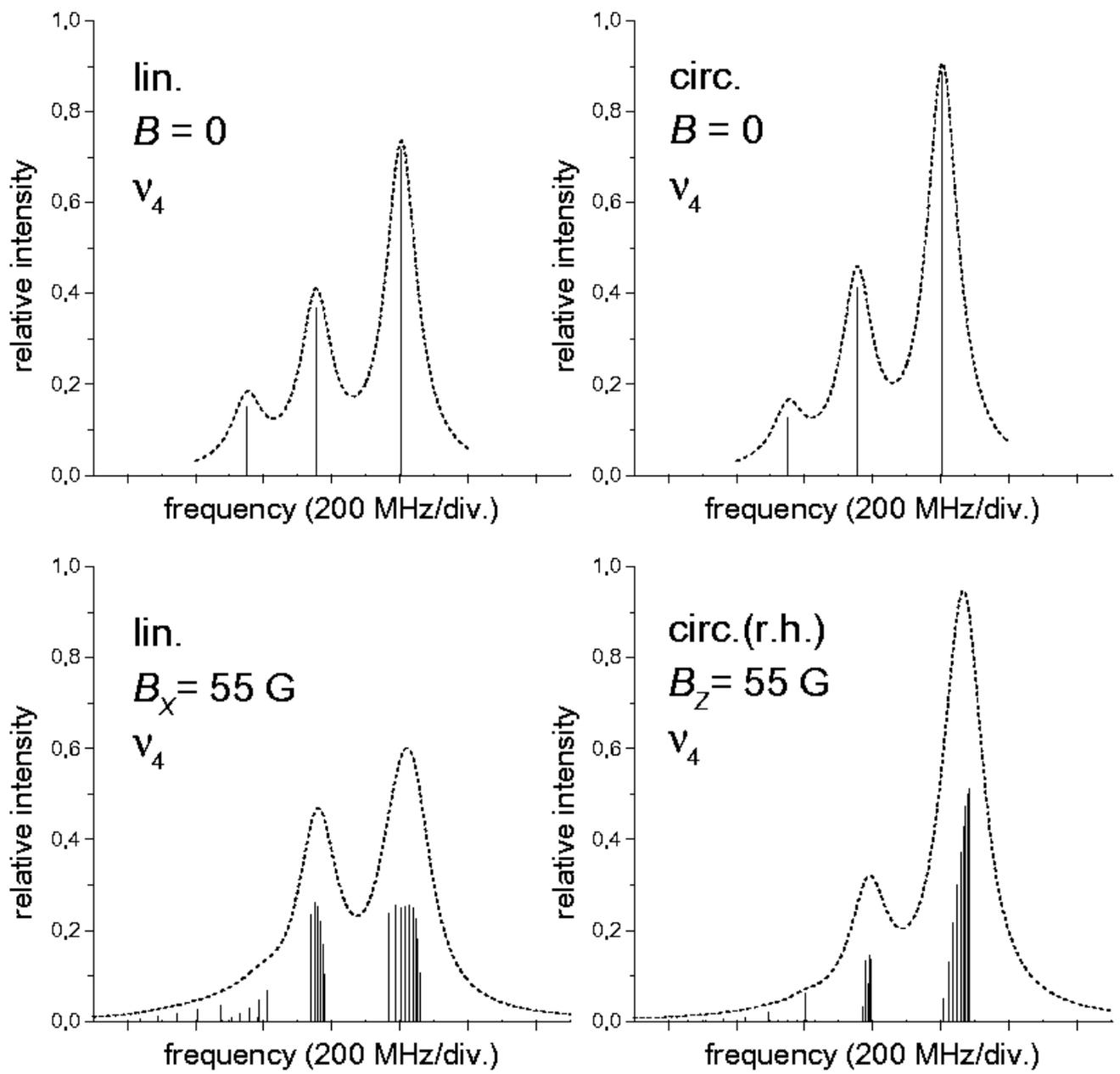
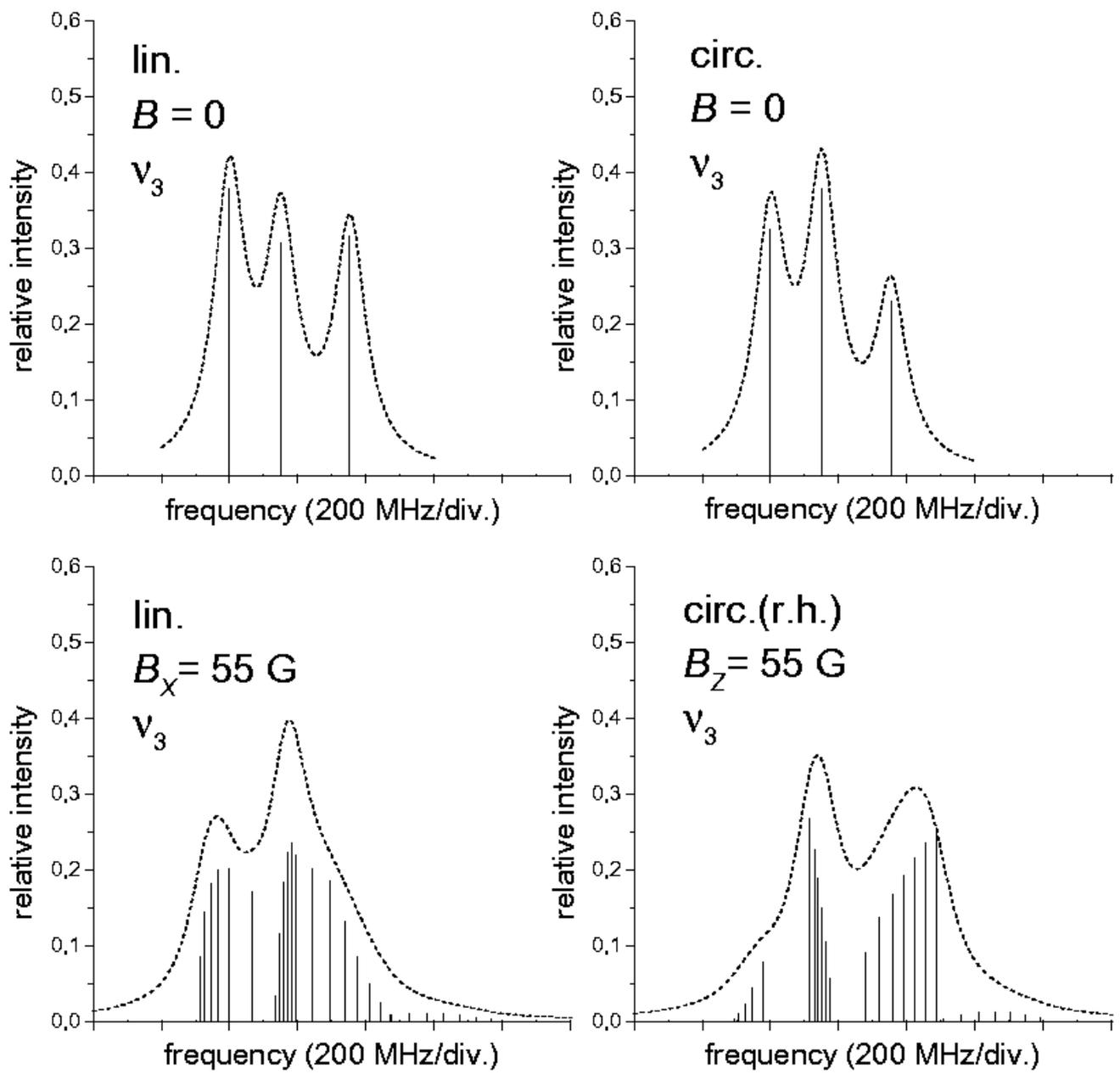
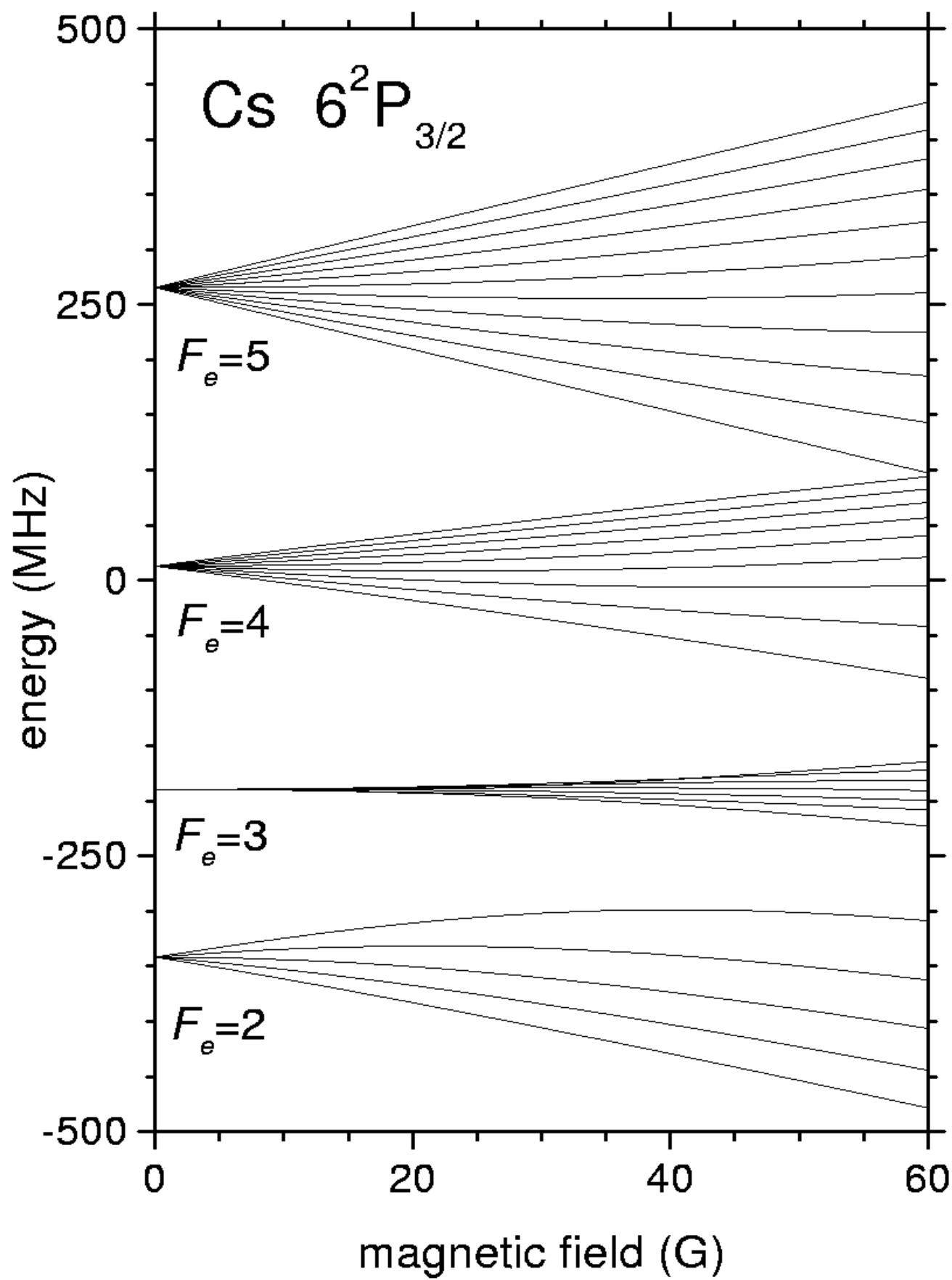
\end{document}